\documentclass[aps,prd,twocolumn,showpacs,nofootinbib]{revtex4-1}

\usepackage{graphicx}
\usepackage{amssymb}
\usepackage{color}

\begin{document}

\title{Renormalization group improved pQCD prediction for $\Upsilon(1S)$ leptonic decay}

\author{Jian-Ming Shen}
\author{Xing-Gang Wu} \email{wuxg@cqu.edu.cn}
\author{Hong-Hao Ma}
\author{Huan-Yu Bi}
\author{Sheng-Quan Wang}

\address{Department of Physics, Chongqing University, Chongqing 401331, P.R. China}
\address{Institute of Theoretical Physics, Chongqing University, Chongqing 401331, P.R. China}

\date{\today}

\begin{abstract}
The complete next-to-next-to-next-to-leading order short-distance and bound-state QCD corrections to $\Upsilon(1S)$ leptonic decay rate $\Gamma(\Upsilon(1S)\to \ell^+\ell^-)$ has been finished by Beneke {\it et al.}~\cite{Beneke:2014qea}. Based on those improvements, we present a renormalization group (RG) improved pQCD prediction for $\Gamma(\Upsilon(1S)\to \ell^+\ell^-)$ by applying the principle of maximum conformality (PMC). The PMC is based on RG-invariance and is designed to solve the pQCD renormalization scheme and scale ambiguities. After applying the PMC, all known-type of $\beta$-terms at all orders, which are controlled by the RG-equation, are resummed to determine optimal renormalization scale for its strong running coupling at each order. We then achieve a more convergent pQCD series, a scheme- independent and more accurate pQCD prediction for $\Upsilon(1S)$ leptonic decay, i.e. $\Gamma_{\Upsilon(1S) \to e^+ e^-}|_{\rm PMC} = 1.270^{+0.137}_{-0.187}$ keV, where the uncertainty is the squared average of the mentioned pQCD errors. This RG-improved pQCD prediction agrees with the experimental measurement within errors.
\end{abstract}


\maketitle

\section{Introduction}

Heavy quarkonium provides an ideal platform for studying the non-relativistic theories, such as the non-relativistic Quantum Chromodynamics (NRQCD)~\cite{Bodwin:1994jh} and the potential NRQCD (PNRQCD)~\cite{Pineda:1997bj, Brambilla:1999xf}. In general, because $v^2_b<v^2_c$ and $\alpha_s(m_b)<\alpha_s(m_c)$, the perturbative results for the bottomonium will be more convergent over the $\alpha_s$- and $v^2$- expansion than the charmonium cases, where $v_{(b,c)}$ stands for the relative velocity of constituent $b$ or $c$ quark in the bottomonium or charmonium rest frame. If enough bottomonium events can be generated at an experimental platform, we can achieve a relatively more definite test of those non-relativistic theories than the charmonium cases.

Being an important high-energy process, the leptonic decay of the ground-state bottomonium $\Upsilon(1S)$ has been studied up to next-to-leading order (NLO)~\cite{Pineda:1996uk, Pineda:2001et}, next-to-next-to-leading order (N$^2$LO)~\cite{Beneke:1999fe, Pineda:2006ri}, and next-to-next-to-next-to-leading order (N$^3$LO)~\cite{Beneke:2014qea}. However, even by including the recently finished complete N$^3$LO pQCD corrections for both the short-distance and the bound-state parts, the pQCD prediction for the decay rate $\Gamma_{\Upsilon(1S)\to e^+ e^-}$ is still about $30\%$ lower than the PDG value, i.e. $\Gamma_{\Upsilon(1S)\to e^+ e^-}|_{\rm Exp.} = 1.340(18)$~keV~\cite{Agashe:2014kda}. Even worse, its pQCD convergence is questionable and one does not know what's the optimal behavior of the running coupling. It is noted that the questionable pQCD series is caused by using conventional scale setting, in which the renormalization scale is simply fixed to be $\sim 3.5$ GeV that leads to maximum decay rate and the renormalization scale uncertainty is predicted by varying it within the range $\mu_{r}\in[3,10]$ GeV~\cite{Beneke:2014qea}. To solve such renormalization scale ambiguity and to improve the pQCD prediction, we shall use the principle of maximum conformality (PMC)~\cite{Brodsky:2011ta, Brodsky:2012rj, Brodsky:2012ik, Mojaza:2012mf, Brodsky:2013vpa, Wu:2014iba} to deal with $\Upsilon(1S)$ leptonic decay rate up to N$^3$LO level.

The PMC provides a systematic procedure to set the optimal renormalization scale for high-energy processes at any order. The behavior of the running coupling is governed by renormalization group (RG)-equation, i.e. the $\beta$-function~\cite{Politzer:1973fx, Politzer:1974fr, Gross:1973id, Gross:1973ju},
\begin{equation}
\beta(a_s)={d a_s(\mu_r)} /{d\ln\mu^2_r}=-a_s^2(\mu_r) \sum_{i=0}^\infty \beta_i a^{i}_s(\mu_r), \label{rge}
\end{equation}
where $a_s=\alpha_s/4\pi$ and $\mu_r$ is the renormalization scale. This provides the underlying principle of PMC, i.e. the optimal behavior of running coupling can be achieved by resumming all the $\{\beta_i\}$-terms of the process that correctly determine the $\alpha_s$-running behavior into the coupling constant. Following the PMC ${\cal R}_\delta$-scheme, the $\beta$-pattern at each perturbative order is a superposition of the $\{\beta_i\}$-terms coming from all the lower-order $\alpha_s$-factors~\cite{Brodsky:2013vpa}. The PMC then resums the $\{\beta_i\}$-series according to the skeleton-like expansion that correctly reproduces the QED limit of the observable~\cite{Brodsky:1997jk}. The resultant PMC scales are functions of the running coupling and are in general different for different orders~\cite{Wu:2013ei}, and the resultant pQCD series is thus identical to a scheme-independent $\beta=0$ conformal series~\cite{Mojaza:2012mf, Brodsky:2013vpa}. After applying the PMC, the pQCD convergence can be generally improved~\footnote{It is noted that there may have $n_f$-terms (ultra-violet free and irrelevant to the $\alpha_s$-renormalization) which should be treated as conformal coefficients~\cite{Ma:2015dxa} and shall not affect our present PMC scale-setting. Their values may be large and may break the pQCD convergence in special cases.}. One reason for such improvement lies in that: Being consistent with the previous treatment in which the $\beta_0$-series are eliminated systematically via the large $\beta_0$-approximation~\cite{Neubert:1994vb, LovettTurner:1994hx, LovettTurner:1995ti, Ball:1995ni, Beneke:1998ui}, the divergent terms $(n!\, \beta^n_i \alpha_s^n)$ disappear in the PMC pQCD series due to the elimination of the RG-$\{\beta_i\}$-terms. It has been found that the PMC follows the RG-invariance and satisfies all the RG-properties~\cite{Brodsky:2012ms}. In the paper, we shall show that after applying the PMC, a more accurate $\Upsilon(1S)$ leptonic decay rate can indeed be achieved.

The remaining parts of the paper are organized as follows. In Sec.II, we will present our calculation technology for the $\Upsilon(1S)$ leptonic decay rate up to N$^3$LO level. In Sec.III, we present numerical results. Sec.IV is reserved for a summary and conclusions. One appendix provide some computational details for PMC.

\section{Calculation technology}

The decay rate for the channel, $\Upsilon(1S)\to \ell^+\ell^-$, can be formulated as
\begin{eqnarray}
\Gamma_{\Upsilon(1S)\to \ell^+\ell^-} = \frac{4\pi \alpha^2}{9 m_b^2} Z_1, \label{eq:master}
\end{eqnarray}
where $\alpha$ is the fine structure constant, $m_b$ is the $b$-quark pole mass, and $Z_1$ stands for the residue of the $1S$-wave two-point correlation function near $(b\bar{b})$-threshold, which can be written as~\cite{Beneke:2007gj}
\begin{eqnarray}
Z_1 = \left|\psi_1(0)\right|^2 c_v \left[ c_v - \frac{E_1}{m_b} \left(c_v+\frac{d_v}{3}\right) + \cdots \right],  \label{eq:z}
\end{eqnarray}
where $c_v$ and $d_v$ are matching coefficients of the leading and sub-leading $(b\bar b)$-currents within the NRQCD framework, whose perturbative forms are
\begin{eqnarray}
c_v = 1+ \sum_{k=1}^n c_k a_s^k, ~~~ d_v = 1+ \sum_{k=1}^n d_k a_s^k, \label{eq:short}
\end{eqnarray}
where $a_s=\alpha_s/4\pi$. Here $|\psi_1(0)|$ and $E_1$ are renormalized wavefunction at the origin and binding energy of $\Upsilon(1{\rm S})$, which represent the bound-state contributions and also receive perturbative corrections from high-order heavy quark potentials and dynamical gluon effect, i.e.
\begin{eqnarray}
E_1 &=& E_1^{(0)}\left(1+\sum_{k=1}^n e_k a_s^k \right), \\
|\psi_1(0)|^2 &=& |\psi_1^{(0)}(0)|^2 \left(1+\sum_{k=1}^n f_k a_s^k\right). \label{eq:bound}
\end{eqnarray}
The LO Coulomb wavefunction at the origin and the LO Coulomb binding energy are given by~\cite{Titard:1993nn, Titard:1994id, Kniehl:1999ud, Melnikov:1998ug, Penin:1998kx}
\begin{eqnarray}
&& \left|\psi_1^{(0)}(0)\right|^2 = \frac{(m_b C_F \alpha_s)^3}{8 \pi}, \\
&& E_1^{(0)} = -\frac{1}{4}m_b (C_F \alpha_s)^2, \label{eq:boundLO}
\end{eqnarray}
where $C_F=4/3$. As a further step, those perturbative coefficients $e_i$ and $f_i$ can be separated as
\begin{eqnarray}
e_i=e_i^{\rm C}+e_i^{\rm nC}+e_i^{\rm us},\;\; f_i=f_i^{\rm C}+f_i^{\rm nC}+f_i^{\rm us},
\end{eqnarray}
where `C', `nC' and `us' denote the corrections from the Coulomb potential, all other non-Coulomb potentials and ultrasoft gluon exchange, respectively. The one-loop and two-loop corrections for the Wilson coefficient $c_v$ have been given by Refs.\cite{Kallen:1955fb, Czarnecki:1997vz, Beneke:1997jm, Kniehl:2006qw}. The fermionic and the purely gluonic three-loop corrections to $c_v$ can be found in Refs.\cite{Marquard:2006qi, Marquard:2009bj, Marquard:2014pea}. The one-loop correction for $d_v$ can be obtained from Ref.\cite{Luke:1997ys}. For the bound state contributions, its NLO term is from the Coulomb potential, and the ultrasoft correction appears first at the third order. Thus, we have $e_1^{\rm nC}=e_1^{\rm us}=0$ and $f_1^{\rm nC}=f_1^{\rm us}=f_2^{\rm us}=0$. The Coulomb, non-Coulomb and ultrasoft corrections to $E_1$ and $|\psi_1(0)|^2$ have been calculated up to N$^3$LO level in Refs.\cite{Beneke:2005hg, Penin:2005eu, Beneke:2007gj, Beneke:2007pj, Beneke:2008cr, Kniehl:2002br}.

Up to N$^n$LO level, one can reformulate the pQCD approximate of the decay rate $\Gamma_{\Upsilon(1S)\to \ell^+\ell^-}$ in a perturbative series as
\begin{eqnarray}
\Gamma_n &=& \sum_{i=0}^{n}\mathcal{C}_{i} \; a_s^{i+3}(\mu_r). \label{eq:expand}
\end{eqnarray}
The LO $\mathcal{C}_0$ can be derived from Eqs.(\ref{eq:master}-\ref{eq:boundLO}), and $\mathcal{C}_i (i\geq1)$ at each order is a combination of the coefficients $c_k$, $d_k$, $e_k$ and $f_k$ at different orders.
There are three energy regions for $\Upsilon(1S)$ leptonic decay, which are characterized by three typical scales, i.e. the hard one $\mu_h\sim m_b$, the soft one $\mu_s\sim m_b v_b$ and the ultra-soft one $\mu_{us} \sim m_b v_b^2$. Because $v_b \sim \alpha_s(m_b v_b)$~\cite{Bodwin:1994jh}, the soft scale $m_b v_b$ is usually replaced by $m_b C_F \alpha_s$, which is the characteristic scale of bottomonium and is connected to its Bohr radius via the relation, $r_{\rm Bohr}=2/(m_b C_F \alpha_s)$.

Practically, one can adopt any value $\mu_r^{\rm init}$ as the initial renormalization scale to do the renormalization, whose value should be large enough to ensure the pQCD calculation. Under the conventional scale setting, i.e. the renormalization scale is fixed to be $\mu_r\equiv\mu_r^{\rm init}$ that is usually choose as the typical momentum of the process, the short-distance and bound-state corrections possess both renormalization and factorization scale ambiguities due to the truncation of perturbative series. The factorization scale problem is another important QCD problem, especially for the present case with several energy scales~\cite{Wu:2013ei}. It has been noted that a proper choice of renormalization scale can lead to a smaller factorization scale dependence~\cite{Wang:2014sua}. In the paper, we shall concentrate our attention on solving the renormalization scale ambiguity and shall take the same choices for factorization scales in different energy regions as those suggested in the literature, that is, we fix the factorization scales as: $\mu_h \equiv m_b$, $\mu_s \equiv C_F \alpha_s(\mu_s) m_b$ and $\mu_{us} \equiv C^2_F \alpha_s^2(\mu_s) m_b$~\cite{Czarnecki:1997vz, Beneke:1997jm, Kniehl:2006qw, Beneke:1997zp, Beneke:1999qg, Beneke:2007gj, Beneke:2005hg, Penin:2005eu, Beneke:2007pj, Beneke:2008cr, Brambilla:2004jw, Anzai:2009tm}.

We note that there exist logarithmic corrections such as the double-logarithmic $\ln^2{\alpha_s}$-terms~\cite{Kniehl:1999mx, Manohar:2000kr} and the single-logarithmic $\ln{\alpha_s}$-terms~\cite{Kniehl:2002yv, Hoang:2003ns} in the perturbative bound-state contributions. The origin of those logarithmic corrections is the presence of several scales in the threshold region. They represent a logarithm of the ratio of scales, e.g. a ratio of the hard scale ($m_b$) to the soft one ($m_b v_b$) or a ratio of the soft one ($m_b v_b$) to the ultra-soft one ($m_b v_b^2$); the resultant $\ln v_b$ equals $\ln{\alpha_s}$ for bound states that are approximately Coulombic, $v_b \propto \alpha_s$~\cite{Kniehl:2002yv}. These corrections are not generated by the renormalization group but are related to the anomalous dimensions of the operators in the effective Hamiltonian~\cite{Pineda:2001et}. Thus in the following PMC treatments, the value of $\ln(\alpha_s)$ is fixed and treated as conformal coefficients, e.g. $\ln(\alpha_s)= \ln(\alpha_s(\mu_s)) \approx-1.1782$.

With all the known results, we are ready to do a PMC analysis of $\Upsilon(1S)$ leptonic decay rate up to N$^3$LO level. The three-loop $\Gamma_3$ can be written as
\begin{eqnarray}
\Gamma_3 &=& c_{1,0} a_s^3(\mu_r^{\rm init})+(c_{2,0}+c_{2,1} n_f)a_s^4(\mu_r^{\rm init}) \nonumber\\
& & +(c_{3,0}+c_{3,1} n_f + c_{3,2} n_f^2) a_s^5(\mu_r^{\rm init})  \nonumber\\
& & +(c_{4,0}+ c_{4,1} n_f + c_{4,2} n_f^2 + c_{4,3} n_f^3)a_s^6(\mu_r^{\rm init}). \label{nfseries}
\end{eqnarray}
The coefficients $c_{i,j}$ $(i>j\geq0)$ at a certain scale can be read from Refs.\cite{Marquard:2014pea, Beneke:2005hg, Penin:2005eu, Beneke:2007gj, Beneke:2007pj, Beneke:2008cr}. In those references, the coefficients are usually given by setting the factorization scales to be equal to the renormalization scale or by directly setting the renormalization scale as $m_b$.

PMC is a kind of $\beta$-resummation, all RG $\{\beta_i\}$-terms should be resummed to form the effective PMC scales. It is thus important to get the correct RG $\{\beta_i\}$-terms of the process. Part of the $\{\beta_i\}$-terms are proportional to the logarithmic terms as $\ln[\mu^{\rm init}_r/\mu_h]$, $\ln[\mu^{\rm init}_r/\mu_s]$, and $\ln[\mu^{\rm init}_r/\mu_{us}]$, which are eliminated by specific choice of renormalization scale in Refs.\cite{Marquard:2014pea, Beneke:2005hg, Penin:2005eu, Beneke:2007gj, Beneke:2007pj, Beneke:2008cr}. Thus before applying the PMC, one should first reconstruct all the coefficients with full factorization and renormalization scale dependence. This goal is achieved by using the scale displacement relation derived from the $\beta$-equation (\ref{rge}), i.e. the coupling $a^k_s(\mu_1)$ at $k_{\rm th}$-order can be related to the coupling at any other scale $\mu_2$ as
\begin{widetext}
\begin{eqnarray}
a^k_s(\mu_1) &=& a^k_s(\mu_2) + k \beta_0 \ln \frac{\mu_2^2}{\mu_1^2} a^{k+1}_s(\mu_2) + k \left(\beta_1 \ln \frac{\mu_2^2}{\mu_1^2} + \frac{k+1}{2}\beta_0^2 \ln^2 \frac{\mu_2^2}{\mu_1^2} \right) a^{k+2}_s(\mu_2) \nonumber\\
&& + k\left[\beta_2 \ln\frac{\mu_2^2}{\mu_1^2}+\frac{2k+3}{2}\beta_0\beta_1 \ln^2 \frac{\mu_2^2}{\mu_1^2}+\frac{(k+1)(k+2)}{3!}\beta_0^3 \ln^3 \frac{\mu_2^2}{\mu_1^2}\right]a^{k+3}_s(\mu_2)+\mathcal{O}[a^{k+4}_s(\mu_2)]. \label{scaledis}
\end{eqnarray}
\end{widetext}
The derived coefficients $c_{i,j}$ $(i>j\geq0)$ with full factorization and renormalization scale dependence are put in the Appendix. As a check of our expressions for $c_{i,j}$, we recover the Eq.(3) of Ref.\cite{Beneke:2014qea} by taking their choices of $\mu_r\equiv\mu_r^{\rm init}$, $\mu_f=\mu_r$  $(\mu_f=\mu_h,\mu_s,\mu_{us})$ and $n_f=4$, and by rewriting $\ln[{\mu_r}/{m_b}]$ as $\ln[\mu_r/(m_b C_F \alpha_s(\mu_r))] +\ln{C_F} +\ln{\alpha_s(\mu_r)}$.

Following the standard PMC procedures as described in detail in Ref.\cite{Brodsky:2013vpa}, we can obtain the required $\{\beta_i\}$-series at each order from Eq.(\ref{nfseries}), i.e.
\begin{widetext}
\begin{eqnarray}
\Gamma_3 &=& r_{1,0}a_s^3(\mu_{r}^{\rm init}) + (r_{2,0}+3\beta_{0}r_{2,1})a_{s}^{4}(\mu_{r}^{\rm init})  +(r_{3,0}+3\beta_{1}r_{2,1}+ 4\beta_{0}r_{3,1}+ 6\beta_{0}^{2}r_{3,2})a_{s}^{5}(\mu_{r}^{\rm init})\nonumber\\
&& +(r_{4,0}+3\beta_{2}r_{2,1}+ 4\beta_{1}r_{3,1} +5\beta_{0}r_{4,1} + \frac{27}{2}\beta_{1}\beta_{0}r_{3,2} +10\beta_{0}^{2}r_{4,2}+10\beta_{0}^{3}r_{4,3}) a_{s}^{6}(\mu_{r}^{\rm init}). ~\label{gammabeta}
\end{eqnarray}
\end{widetext}
The $\beta_i$-coefficients $r_{i,j}$ $(i>j\geq0)$ can be obtained from the $n_f$-coefficients $c_{i,j}$ $(i>j\geq0)$ by applying basic PMC formulas listed in Ref.\cite{Brodsky:2013vpa}. The non-conformal coefficients $r_{i,j}$ $(j\neq0)$ are functions of $\mu_r^{\rm init}$; while, the conformal coefficients $r_{i,0}$ are independent of $\mu_r^{\rm init}$. For convenience, we present the conformal coefficients $r_{i,0}$ with explicit factorization scale and/or initial scale dependence in the Appendix.

After applying the PMC, the three-loop leptonic decay rate $\Upsilon(1S)$ changes to
\begin{displaymath}
\Gamma_3 = r_{1,0}a_s^3(Q_1) + r_{2,0}a_s^4(Q_2)+r_{3,0}a_s^5(Q_3)+ r_{4,0}a_s^6(Q_4),
\end{displaymath}
where $Q_i (i=1,2,3,4)$ are PMC scales at each perturbative order, whose expressions with explicit factorization scale and/or initial scale dependence are put in the Appendix. To eliminate the non-conformal $\beta$-terms, the renormalization scales at each perturbative order have been shifted from its initial value $\mu_r^{\rm init}$ to the optimal ones $Q_i$ at different orders. The PMC scales at each order are determined unambiguously by resuming all the same type of non-conformal $\beta$-terms governed by RG-equation into the running coupling. The resulting pQCD series is identical to the one of the conformal theory with $\beta=0$ and is thus scheme independent. The PMC scales correctly characterize the virtuality of the propagating gluons and thus also allow one to determine the value of the effective number of flavors $n_f$. For the present decay process, the number of active flavors is fixed by the number of quarks in the effective theory. Since the bottom and the top quark have been integrated out, thus for self-consistency, we shall fix $n_f=4$ and adopt the four-flavor $\alpha_s$-running to do our discussions. Because of lacking even higher-order $\{\beta_i\}$-terms, we cannot determine $Q_4$, and we simply set $Q_4=Q_3$ in the following calculation. This treatment will lead to residual scale dependence, which, however, will be highly suppressed~\cite{Wu:2013ei}.

\section{Numerical results}

We adopt $N_c=3$ for the $SU(N_c)$-color group and adopt the four-loop $\alpha_s$-running to do the numerical analysis of the $\Upsilon(1S)$ leptonic decay rate up to three-loop QCD corrections. By taking $\alpha_s(M_{Z})= 0.1185$~\cite{Agashe:2014kda}, we obtain $\Lambda^{(n_f=4)}_{\rm QCD}=0.301$ GeV. We take the fine structure constant $\alpha(2m_b)=1/132.3$~\cite{Jegerlehner:2011mw}. Using the highest known three-loop relation between the pole mass and $\overline{\rm MS}$-running mass and taking the $b$-quark $\overline{\rm MS}$-mass $\bar{m}_b(\bar{m}_b)= 4.180$ GeV~\cite{Agashe:2014kda}, we obtain the $b$-quark pole mass $m_b=4.922$ GeV \footnote{The choice of $b$-quark pole mass and also $|\psi_1^{(0)}(0)|^2$ and $E^{(0)}_1$ in final expresses ensure the correct using of PMC, since only those $\beta$-terms that are pertained to the renormalization of the running coupling should be absorbed into the running coupling. Here we also do not consider the non-perturbative corrections/uncertainties for $|\psi^{(0)}_{1}(0)|$ and $E^{(0)}_1$. }.

\begin{figure}[htb]
\includegraphics[width=0.50\textwidth]{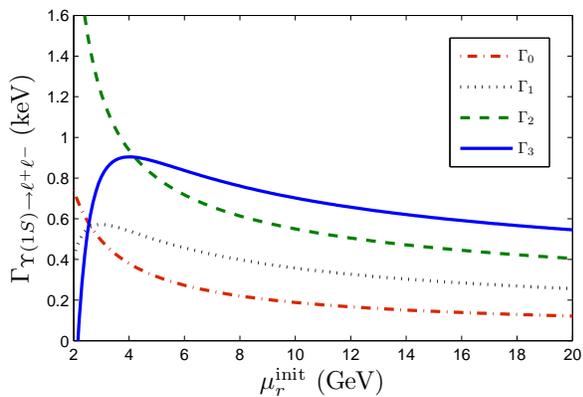}
\caption{The decay rate $\Gamma_n$ with $n=(0,1,2,3)$ under the conventional sale setting method as a function of the initial choice of renormalization scale $\mu^{\rm init}_r$, where $\Gamma_n$ is defined by Eq.(\ref{eq:expand}) and stands for the decay rate with up to $n_{\rm th}$-loop QCD corrections. }  \label{conv}
\end{figure}

We first present the decay rate $\Gamma_{\Upsilon(1S)\to \ell^+\ell^-}$ with different loop corrections in Fig.(\ref{conv}), in which the conventional scale setting method with the renormalization scale $\mu_r\equiv \mu^{\rm init}_r$ is adopted. To be self-consistent, when calculating $\Gamma_n$, the $(n+1)_{\rm th}$-loop $\alpha_s$-running together with its own $\Lambda_{\rm QCD}$ value are adopted. Fig.(\ref{conv}) agrees with the conventional wisdom that with the increment of loop corrections, the conventional scale dependence becomes smaller. It also indicates that the higher-order terms are important for an accurate pQCD prediction.

\begin{figure}[htb]
\includegraphics[width=0.50\textwidth]{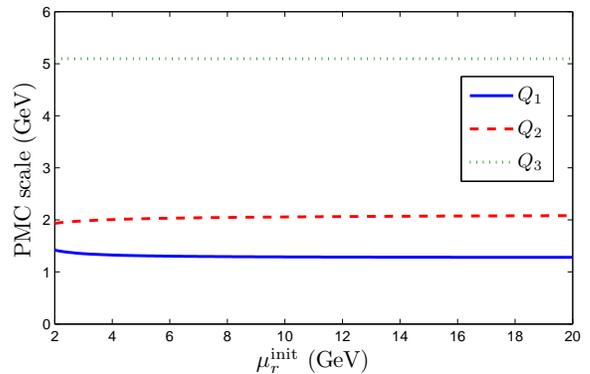}
\caption{The PMC scales $Q_i$ with $i=(1,2,3)$ at each perturbative order versus the initial renormalization scale $\mu_r^{\rm init}$. The solid, the dashed, and the dotted lines are for $Q_1$, $Q_2$, and $Q_3$, respectively. }  \label{fig:scale}
\end{figure}

In Fig.(\ref{fig:scale}), we present the initial scale dependence for the PMC scales $Q_1$, $Q_2$ and $Q_3$. Fig.(\ref{fig:scale}) shows that the PMC scales $Q_i$ are almost independent on the choice of initial renormalization scale $\mu_r^{\rm init}$ by varying it within a large perturbative region such as $2\sim20$ GeV. If setting $\mu_r^{\rm init}=m_b$, we find the LO PMC scale $Q_1\simeq1.31$ GeV, the NLO PMC scale $Q_2\simeq2.02$ GeV and the N$^2$LO PMC scale $Q_3\simeq5.10$ GeV. Those scales are different from the guessed value $\sim 3.5$ GeV that leads to maximum decay rate under conventional scale setting.

\begin{table}[htb]
\centering
\begin{tabular}{cccccc}
\hline
      & ~~LO~~    & ~~NLO~~    & ~~N$^2$LO~~ & ~~N$^3$LO~~ & ~~sum~~\\
\hline
Conv. & ~~+0.374~~ & ~~+0.125~~ & ~~+0.322~~   & ~~+0.061~~  & +0.882 \\
PMC   & ~~+2.292~~ & ~~-1.198~~ & ~~+0.191~~   & ~~-0.015~~  & +1.270 \\
\hline
\end{tabular}
\caption{Contributions from each order for the three-loop decay rate $\Gamma_3$ (in unit: keV) under the conventional (Conv.) and the PMC scale settings, respectively. $\mu^{\rm init}_{r}=m_b$. }  \label{tab:eachorder}
\end{table}

\begin{table}[htb]
\centering
\begin{tabular}{cccc}
\hline
      & $K_1$ & $K_2$ & $K_3$ \\
\hline
Conv. & ~33.3\%~ & ~64.7\%~  & ~7.4\%~  \\
PMC   & ~52.3\%~ & ~17.5\%~  & ~1.1\%~  \\
\hline
\end{tabular}
\caption{The defined $K$ factor ($K_n$) for the N$^n$LO term of $\Gamma_3$ before and after the PMC scale setting, where $n=1$, $2$ and $3$, respectively. $\mu^{\rm init}_{r}=m_b$. }  \label{tab:kfactor}
\end{table}

The non-conformal terms determine the renormalization scales at each perturbative order and the conformal terms as well as the resultant PMC scales accurately display the magnitude of the pQCD correction at each perturbative order. We present the contributions from each order for $\Gamma_3$ in Table~\ref{tab:eachorder}, in which the results before and after the PMC scale setting are presented. Under conventional scale setting, the N$^2$LO term is about $90\%$ of the LO term, and is almost three times of the NLO term, breaking the pQCD nature of the series. After applying the PMC, the pQCD convergence is improved: the magnitude of N$^2$LO term is about $16\%$ of the NLO term and the magnitude of N$^3$LO term is about $8\%$ of the N$^2$LO term. This can be show more clearly by defining a $K$ factor ($K_n$) that equals to the magnitude of the ratio between the $n_{\rm th}$-order term and the sum of all lower-order terms. The $K$ factors for NLO, N$^2$LO and N$^3$LO terms are presented in Table \ref{tab:kfactor}.

\begin{figure}[htb]
\includegraphics[width=0.50\textwidth]{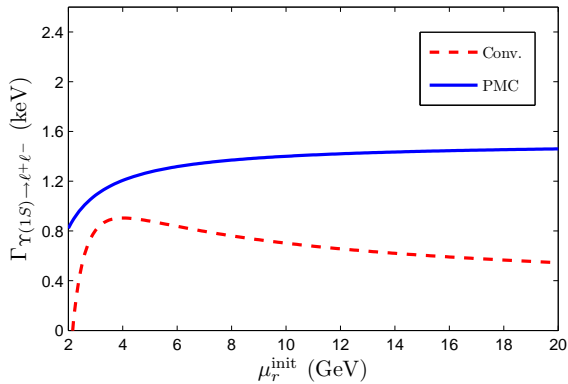}
\caption{The $\Gamma_3$ versus the initial renormalization scale $\mu_r^{\rm init}$ before and after the PMC scale setting. The dashed and solid lines are for the conventional (Conv.) and the PMC scale settings, respectively. }  \label{fig:gamma2}
\end{figure}

In Fig.(\ref{fig:gamma2}), we present the three-loop $\Gamma_3$ versus the choice of initial scale $\mu^{\rm init}_r$, in which the results before and after the PMC scale setting are presented as a comparison. Under conventional scale setting, the decay rate $\Gamma_3$ shall first increase and then decrease with the increment of $\mu^{\rm init}_r$; If setting $\mu_r\equiv\mu^{\rm init}_r\sim 3.5$ GeV, we obtain its maximum value, which however is still lower than the central PDG value by about $30\%$. After applying the PMC, the decay rate $\Gamma_3$ monotonously raises with the increment of $\mu^{\rm init}_r$, and the renormalization scale dependence has been greatly suppressed. By taking a hard enough scale such as $\mu_r^{\rm init}>4$ GeV, the computed PMC scales and the final PMC prediction for the leptonic $\Upsilon(1S)$ decay are highly independent to its exact values. If taking $\mu_r^{\rm init}=m_b$, we obtain
\begin{equation}
\Gamma_{\Upsilon(1S)\to \ell^+ \ell^-} = 1.270 ~{\rm keV}, \label{eq:pmcc}
\end{equation}
which is consistent with the central PDG value within $5\%$ error~\cite{Agashe:2014kda}. In Ref.~\cite{Pineda:2006ri} the authors achieved a better NNLO prediction by including full resummation of logarithms at next-to-leading-logarithmic accuracy and partial contributions at next-to-next-to-leading logarithmic accuracy. The improvement of the pQCD convergence and scale dependence is in some sense consistent with the PMC prediction. This can be explained by the fact that the large logarithmic terms are usually accompanied by certain $\{\beta_i\}$-terms, thus the resummation of large log-terms could be consistent with the PMC $\beta$-resummation.

For the present process, the perturbative series starts at $\alpha^3_s$-order, slight change of its argument shall result in large pQCD error, thus this process provides a good platform for testing the correct running behavior of the coupling constant. On the one hand, the PMC prediction for $\Upsilon(1S)$ leptonic decay reads
\begin{eqnarray}
\Gamma_{\Upsilon(1S) \to \ell^+ \ell^-}|_{\rm PMC} &=& 1.270 ^{+0.130 +0.043}_{-0.182 -0.042} \pm 0.015  ~{\rm keV}  \\
&=& 1.270^{+0.137}_{-0.187} ~{\rm keV},
\end{eqnarray}
where the first error is the residual initial scale dependence for $\mu_r^{\rm init} \in[3,10]\;{\rm GeV}$, the second error is for $\alpha_s(M_{Z})_{\rm Exp.}= 0.1185\pm0.0006$~\cite{Agashe:2014kda}, and the third error is the estimated unknown high-order contributions. The errors in the second line stand for the squared averages of those errors. The unknown high-order contribution is predicted as $\pm |\mathcal{C}_3 a_s^{6}|_{\rm MAX}$~\cite{Wu:2014iba}, where the symbol ``MAX" stands for the maximum $|\mathcal{C}_3 a_s^{6}|$ within the region of $\mu_r^{\rm init} \in[3,10]\;{\rm GeV}$. This RG-improved pQCD prediction agrees well with the experimental measurement. It is noted that for the present case, even though the PMC scales themselves are almost flat within the region of $\mu_r^{\rm init} \in[3,10]\;{\rm GeV}$, cf. Fig.(\ref{fig:scale}), there is large residual scale dependence in comparison to the previous PMC examples, such as Refs.\cite{Wang:2013akk, Wang:2014wua, Wang:2014aqa, Wang:2013bla}. Thus we need to know even higher-order $\beta$-terms for this particular process so as to achieve accurate PMC scales and PMC predictions.

On the other hand, the present PMC prediction on the $\Upsilon(1S)$ decay rate together with its errors can be compared with the prediction under the conventional scale setting
\begin{eqnarray}
\Gamma_{\Upsilon(1S)\to \ell^+ \ell^-}|_{\rm Conv.} &=& 0.882 ^{+0.022+0.023}_{-0.180-0.022} \pm0.443 ~{\rm keV}   \\
&=& 0.882 ^{+0.444}_{-0.479} ~{\rm keV},
\end{eqnarray}
where the first error is initial scale dependence for $\mu_r^{\rm init} \in[3,10]\;{\rm GeV}$, the second error is from $\alpha_s(M_{Z})|_{\rm Exp.}$ uncertainty, and the third error is the estimated unknown higher-order contributions. The errors in the second line stand for the squared averages of those errors. The central decay rate is lower than the central PDG value by about $34\%$, and the much larger errors in comparison to the PMC prediction are caused by the large value of N$^3$LO term at the scale $3$ GeV, which are consistent with observation shown in Ref.\cite{Beneke:2014qea}.

\begin{table}[htb]
\centering
\begin{tabular}{cccc}
\hline
      & ~~$\mu_h$~~    & ~~$\mu_s$~~    & ~~$\mu_{us}$~~ \\
\hline
      & ~~$+0.004$~~ & ~~$+0.039$~~ & ~~$+0.002$~~ \\
\raisebox {2.0ex}[0pt]{~~$\Delta\Gamma_{3}|_{\mathrm{Conv.}}$~~}
      & ~~$-0.004$~~ & ~~$-0.033$~~ & ~~$-0.003$~~ \\
\hline
      & ~~$+0.003$~~ & ~~$+0.120$~~ & ~~$+0.002$~~ \\
\raisebox {2.0ex}[0pt]{~~$\Delta\Gamma_{3}|_{\mathrm{PMC}}$~~}
      & ~~$-0.004$~~ & ~~$-0.091$~~ & ~~$-0.003$~~ \\
\hline
\end{tabular}
\caption{The factorization uncertainties $\Delta\Gamma_3$ (in units of keV) for the three-loop decay rate $\Gamma_3$ before and after the PMC scale setting, which are caused by separately varying $\mu_h$, $\mu_s$ and $\mu_{us}$ by $\pm10\%$ of their center values, respectively. }  \label{tab:factError}
\end{table}

\begin{figure}[htb]
\includegraphics[width=0.50\textwidth]{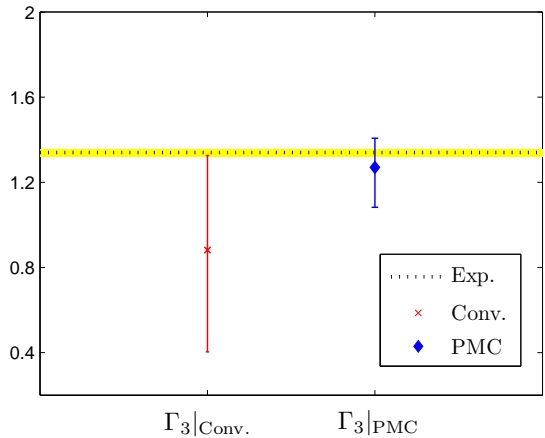}
\caption{A comparison of $\Gamma_3$ together with its pQCD errors before and after the PMC scale setting. The theoretical errors are squared average of all the mentioned uncertainties. The PDG value, $\Gamma_{\Upsilon(1S)\to e^+ e^-}|_{\rm Exp.} = 1.340(18)$~keV~\cite{Agashe:2014kda}, is included as a comparison. }  \label{fig:gamma3}
\end{figure}

Let us end with a final comment on the factorization scale dependence. At present, we have no strict and systematic way to set the factorization scale, and the question is much more involved when there are several scale regions. As a reference, we present a discussion of factorization scale uncertainties under several simple choices of factorization scales, whose values before and after the PMC sale setting are presented in Table.~\ref{tab:factError}. Here, to ensure the effectiveness of the NRQCD and pNRQCD factorization approaches, we vary the scales $\mu_h$, $\mu_s$ and $\mu_{us}$ separately by $\pm10\%$ of their center values; e.g. when discussing the uncertainty of $\mu_h$, we take $\mu_h=(1\pm10\%)m_b$ and fix $\mu_s$ and $\mu_{us}$ to be their central values; the uncertainties for $\mu_s$ and $\mu_{us}$ are done via the same way. Table.~\ref{tab:factError} shows that after applying the PMC, the factorization scale uncertainties are still there and the largest uncertainty is caused by the soft scale $\mu_s$. As shown by Eqs.(\ref{fac1},\ref{fac2},\ref{fac3}), the PMC scales depend on the factorization scales. More explicitly, when setting $\mu_s = 90\% C_F \alpha_s(\mu_s) m_b$, the value of $Q_1$ changes from $1.31 \mathrm{GeV} \to 1.27 \mathrm{GeV}$, the value of $Q_2$ changes from $2.02 \mathrm{GeV} \to 1.96\mathrm{GeV}$, and the value of $Q_3$ changes from $5.10 \mathrm{GeV} \to 4.78 \mathrm{GeV}$. Those are slight scale changes, however they shall lead to sizable contributions, since the decay rate starts at $\alpha_s^3$-order.

\section{summary}

We have studied the N$^3$LO short-distance and bound-state QCD corrections to $\Upsilon(1S)$ leptonic decay rate of $\Upsilon(1S)\to \ell^+\ell^-$ by applying the PMC. A comparison of the three-loop $\Gamma_3$ together with its pQCD errors before and after the PMC scale setting is presented in Fig.(\ref{fig:gamma3}), where the theoretical errors are squared average of all the mentioned pQCD uncertainties. It shows that our present RG-improved pQCD prediction agrees well with the experimental measurement within errors. After applying the PMC, the pQCD convergence of the resultant series is improved. Thus, the PMC does provide a systematic and unambiguous way to set the renormalization scale for any QCD processes and the accuracy of the pQCD prediction can be greatly improved. It is noted that we have not considered the non-perturbative corrections/uncertainties for $|\psi_{1}(0)|$ and $E_1$, and for the decay rate $\Gamma(\Upsilon(1S)\to \ell^+\ell^-)$. Those studies shall further improve our present PMC predictions, which are out of the range of the present paper.

\hspace{0.5cm}

\noindent {\bf Acknowledgement}: This work was supported in part by Natural Science Foundation of China under Grant No.11275280, and by Fundamental Research Funds for the Central Universities under Grant No.CDJZR305513.  \\

\appendix

\section{The coefficients $c_{i,j}$, the conformal coefficients $r_{i,0}$ and the PMC scales for $\Gamma_3$}

As mentioned in the body of the text, before applying the PMC scale setting, one should reconstruct all the coefficients with full factorization and renormalization scale dependence. In this Appendix, we first present the coefficients $c_{i,j}$ for $n_f$-power series, and then present conformal coefficients $r_{i,0}$ and the PMC scales $Q_{i}$ for the three-loop $\Upsilon(1S)$ leptonic decay rate $\Gamma_3$. The full renormalization scale and factorization scale dependence shall be explicitly presented.

We take the two-loop coefficients $c_{3,j}$ as an example to explain the reconstruction procedures. As this perturbative order, we need to deal with the two-loop QCD corrections to both $c_v$ and $|\psi_1(0)|$. We fix the scale dependence for $c_v$ and $|\psi_1(0)|$ separately, which are in different energy regions.

The expression of the two-loop $c_v(\mu_r=m_b,\mu_f=\mu_h)$ can be found in Refs.\cite{Czarnecki:1997vz, Beneke:1997jm}, which involves only one factorization scale $\mu_h$. The expression for $c_v(\mu_r,\mu_h)$ at arbitrary choice of $\mu_r$ can be derived by replacing $\alpha_s(m_b)$ in $c_v(m_b,\mu_h)$ to $\alpha_s(\mu_r)$ with the help of the scale displacement relation (\ref{scaledis}). By setting $\mu_1=m_b$, $\mu_2=\mu_r$ and $k=1$ in Eq.(\ref{scaledis}), we obtain the required full factorization and renormalization scale dependence for $c_v(\mu_r,\mu_f=\mu_h)$, which is
\begin{widetext}
\begin{eqnarray}
c_v(\mu_r,\mu_h) &=& 1 - 2C_F \frac{\alpha_s(\mu_r)}{\pi} + \left(\frac{\alpha_s(\mu_r)}{\pi}\right)^2\bigg[{1\over2}C_F {\beta_0}\ln\frac{\mu_r^2}{m_b^2} + \left({22\over9}-{{2\pi^2}\over9}\right) C_F T_F n_h -
\left( {1\over4}C_A + {1\over6}C_F \right) \pi^2 C_F \ln\frac{\mu_h^2}{m_b^2} \nonumber\\
&& +{11\over18}C_F T_F n_f - \left({{89\pi^2}\over144}-{5\over6}\pi^2\ln2-{{13\zeta(3)}\over4}-{151\over72}\right)C_A C_F
+ \left({23\over8} - \frac{79\pi^2}{36} + \pi^2\ln2 - \frac{\zeta(3)}{2}\right)C_F^2\bigg],   \label{eq:cv2loop}
\end{eqnarray}
\end{widetext}
where $C_A$, $C_F$ and $T_F$ are quadratic Casimir invariants~\cite{vanRitbergen:1998pn}. For a $SU(N_c)$-color group, we have $C_A=N_c$, $C_F=(N^2_c -1) / 2 N_c$ and $T_F=1/2$. In the present case, $n_h=1$, however, we keep $n_h$ in Eq.(\ref{eq:cv2loop}) for convenience.

The expression of the two-loop $|\psi_1(0)|^2$ can be found in Ref.\cite{Beneke:2005hg}, in which the renormalization scale $\mu_r$ is set to be the soft scale $\mu_s$. To get the expression for $\mu_r\neq \mu_s$, we can replace $\alpha_s(\mu_s)$ by $\alpha_s(\mu_r)$ with the help of Eq.(\ref{scaledis}); e.g. by setting $\mu_1=\mu_s$, $\mu_2=\mu_r$ and $k=1$ in Eq.(\ref{scaledis}), we obtain
\begin{widetext}
\begin{eqnarray}
|\psi_1^{(2)}(0)|^2 &=& |\psi_1^{(0)}(0)|^2 \bigg\{1+ \left(6 \beta _0 L(\mu_r)+c_{\psi,1}^{\rm C}\right) a_s(\mu_r) + \bigg[(36+16 L(\mu_h))\pi^2 C_A C_F+\left(\frac{178}{9}+\frac{32 L(\mu_h)}{3}\right)\pi^2 C_F^2
\nonumber\\
&& +{\beta_0^2}\left(24L^2(\mu_r)-12L(\mu_s)\right) +6{\beta_1}L(\mu_r)+ 8{\beta_0}{c_{\psi,1}^{\rm C}}L(\mu_r) +c_{\psi,2}^{\rm C} \bigg] a_s^2(\mu_r) \bigg\},
\end{eqnarray}
\end{widetext}
where $\mu_h$ and $\mu_s$ stand for the hard and the soft scales, respectively. $c_{\psi,1}^{\rm C}$ and $c_{\psi,2}^{\rm C}$ are non-logarithmic parts of the first-order and the second-order Coulomb corrections~\cite{Melnikov:1998ug, Beneke:1999qg}, and for the present $1S$-wave bound state, they can be simplified as
\begin{eqnarray}
c_{\psi,1}^{\rm C} &=& 2.6229-1.61351{n_f}, \nonumber\\
c_{\psi,2}^{\rm C} &=& 1800.75-193.489{n_f}+3.50376{n_f^2}. \nonumber
\end{eqnarray}
The logarithmic function $L(x)$ is defined as
\begin{eqnarray}
L(x)=\ln[x/(m_b C_F \alpha_s(\mu_s))],
\end{eqnarray}
where $x=\mu_r$, $\mu_h$ or $\mu_s$, respectively.

As a combination, we get the required coefficients $c_{3,j}$. Following the similar treatment, we can derive the expressions for all the coefficients $c_{i,j}$ $(i>j\geq0)$ with full scale dependence up to three-loop level from the ones at a particular scale presented in Refs.\cite{Marquard:2014pea, Beneke:2005hg, Penin:2005eu, Beneke:2007gj, Beneke:2007pj, Beneke:2008cr, Kniehl:2002br}.

In using the original results of Refs.\cite{Beneke:2005hg, Beneke:2007gj, Beneke:2007pj}, there are some subtleties in deriving the full scale dependence of the three-loop coefficient $f_3=f_3^{\rm C}+f_3^{\rm nC}+f_3^{\rm us}$ for $|\psi_1(0)|^2$. Most of logarithmic terms for the Coulomb correction $f_3^{\rm C}$ given in Ref.\cite{Beneke:2005hg} are for $\mu_r= \mu_s$, and there is one logarithmic term that has explicit ultrasoft scale dependence, which originates from the non-Abelian gluon ``H-diagram''~\cite{Brambilla:2004jw, Anzai:2009tm} and should be treated as conformal coefficients. The logarithmic terms for the non-Coulomb correction $f_3^{\rm nC}$ should be rewritten as $\ln\mu^2_h/m^2_b$ and $\ln\mu^2_s/(m_b C_F \alpha_s(\mu_s))^2$~\cite{Beneke:2007gj, Kniehl:2002br}. The logarithmic terms for the ultrasoft correction $f_3^{\rm us}$ are for $\mu_r= \mu_s$~\cite{Beneke:2007pj, Beneke:2008cr}. After this clarification, we are ready to derive the full scale dependence of $f_3$ with the help of Eq.(\ref{scaledis}), which reads
\begin{widetext}
\begin{eqnarray}
|\psi_1^{(3)}(0)|^2 &=& |\psi_1^{(2)}(0)|^2 + |\psi_1^{(0)}(0)|^2 a_s^3(\mu_r) \bigg\{ \pi^2 C_A^3 \bigg[ 32-\frac{16\pi^2}{3} -48\ln{\frac{\mu_s}{m_b}}+64 \ln{(\alpha_s(\mu_s))}+48 L(\mu_{us})\bigg] \nonumber\\
&& +\pi^2 C_F C_A^2 \bigg[ \bigg(\frac{320}{3}\ln{2} -\frac{704}{3} +\frac{512}{3} {H_1} \bigg)\ln{\frac{\mu_s}{m_b}}-\frac{160}{3}\ln^2{\frac{\mu_s}{m_b}} +\frac{512}{3} \ln{(\alpha_s(\mu_s))} \ln{\frac{\mu_s}{m_b}}
\nonumber\\
&& +\bigg( \frac{3328}{9}-256{H_1}-\frac{512}{3}\ln{2} \bigg) \ln{(\alpha_s(\mu_s))}-128\ln^2{(\alpha_s(\mu_s))} +\bigg(\frac{6176}{9}+\frac{256}{3}\ln{2}\bigg)L(\mu_s)
\nonumber\\
&& +\frac{256}{3}L^2(\mu_s) \bigg] + {\pi^2 C_A C_F^2}\bigg[ 8\ln^2{\frac{\mu_h}{m_b}} -\frac{32}{9}\ln{\frac{\mu_h}{m_b}}
+32L(\mu_s)\ln {\frac{\mu_h}{m_b}} +\bigg( \frac{22240}{27}+\frac{512}{3}\ln{2} \bigg) L(\mu_s)
\nonumber\\
&& +\frac{592}{3}L^2(\mu_s) +\bigg( \frac{32}{3}+\frac{1280}{3}{H_1}+\frac{64}{3}\ln{2} \bigg) \ln{\frac{\mu_s}{m_b}} -\frac{352}{3}\ln^2{\frac{\mu_s}{m_b}} +\frac{1280}{3} \ln{\frac{\mu_s}{m_b}} \ln{(\alpha_s(\mu_s))}
\nonumber\\
&& +\bigg( \frac{512}{3}-\frac{256}{3}\ln{2}-\frac{2048}{3}{H_1} \bigg)\ln{(\alpha_s(\mu_s))}
 -\frac{1024}{3} \ln^2{(\alpha_s(\mu_s))}\bigg] + \pi^2{C_A C_F} \bigg[-\frac{1744}{9}T_F n_f L(\mu_s)
\nonumber\\
&& +\bigg( 128L^2(\mu_s)-160L(\mu_h)L(\mu_s) +\bigg(16-\frac{32}{3}\pi^2\bigg)L(\mu_s) +360L(\mu_r)+160L(\mu_r)L(\mu_h) \bigg) \beta_0 \bigg]
\nonumber\\
&& +{\pi^2 C_F^3} \bigg[ \frac{80}{3}\ln{\frac{\mu_h}{m_b}} +\frac{16}{3}\ln^2{\frac{\mu_h}{m_b}} +\frac{64}{3}L(\mu_s)\ln{\frac{\mu_h}{m_b}} +\frac{160}{3}L(\mu_s) +\frac{224}{3}L^2(\mu_s) -\frac{512}{3}\ln^2{(\alpha_s(\mu_s))}
\nonumber\\
&& +\bigg(512\ln{2}-\frac{4096}{9}-\frac{1024}{3}{H_1}\bigg)\ln{(\alpha_s(\mu_s))} +\bigg(\frac{1088}{3}+\frac{512}{3}{H_1}-384\ln{2}\bigg)\ln{\frac{\mu_s}{m_b}} -\frac{64}{3}\ln^2{\frac{\mu_s}{m_b}}
\nonumber\\
&&  +\frac{512}{3}\ln{\frac{\mu_s}{m_b}}\ln{(\alpha_s(\mu_s))} \bigg] +{\pi^2 C_F^2 T_F}\bigg(\frac{64}{15}\ln{\frac{\mu_h}{m_b}} +\frac{128}{15}L(\mu_s) -\frac{3776}{27}{n_f}L(\mu_s)\bigg) +{\pi^2 C_F^2 \beta_0}
\nonumber\\
&& \times\bigg[ \bigg(\frac{128}{9}-\frac{64}{9}{\pi^2}\bigg)L(\mu_s) +\frac{256}{3}L^2(\mu_s) -\frac{320}{3}L(\mu_h)L(\mu_s) +\frac{320}{3}L(\mu_r)L(\mu_h) +\frac{1780}{9}L(\mu_r) \bigg]
\nonumber\\
&& +10{\beta_0}{c_{\psi,2}^{\rm C}}L(\mu_r) +8{\beta_1}{c_{\psi,1}^{\rm C}}L(\mu_r) +8{\beta_0^2}{c_{\psi,1}^{\rm C}}\bigg(5L^2(\mu_r)-2L(\mu_s)\bigg) +6{\beta_0}{\beta_1}\bigg(9L^2(\mu_r)-4L(\mu_s)\bigg)
\nonumber\\
&& +6{\beta_2}L(\mu_r) +4{\beta_0^3}\bigg(6L(\mu_s)+3L^2(\mu_s)-30L(\mu_s)L(\mu_r)+20L^3(\mu_r)\bigg) +c_{\psi,3}^{\rm C} + c_{\psi,3}^{\rm nC} +c_{\psi,3}^{\rm us}  \bigg\}.
\end{eqnarray}
\end{widetext}
where $H_1=\ln{C_F}-1$, and for the $1S$-wave bound state the non-logarithmic part of the third-order Coulomb, non-Coulomb and ultrasoft corrections~\cite{Beneke:2005hg,Beneke:2007gj,Beneke:2007pj} can be simplified as,
\begin{eqnarray}
c_{\psi,3}^{\rm C} &=& -39854.2+2005.08{n_f}+19.7985{n_f^2}+3.61806{n_f^3}, \nonumber\\
c_{\psi,3}^{\rm nC} &=& -44754.7-3126.52{n_f}, \nonumber\\
c_{\psi,3}^{\rm us} &=& 223012. \nonumber
\end{eqnarray}

All the coefficients under the arbitrary choice of initial renormalization scale $\mu^{\rm init}_r$ that are adopted in the body of text are in the following (in unit of GeV):
\begin{widetext}
\begin{eqnarray}
c_{1,0}&=& 0.0734844, \\
c_{2,0}&=& -1.37493+4.84997L(\mu_r^{\rm init}), \\
c_{2,1}&=& -0.118568-0.293938L(\mu_r^{\rm init}), \\
c_{3,0}&=& 168.629-41.5323L(\mu_r^{\rm init})+213.399L^2(\mu_r^{\rm init})+60.1699L(\mu_h)-106.699L(\mu_s) \nonumber\\
&& -34.4887\ln{\frac{\mu_r^{\rm init}}{m_b}}-60.1699\ln{\frac{\mu_h}{m_b}}, \\
c_{3,1}&=& -10.7309-10.7761L(\mu_r^{\rm init})-25.8665L^2(\mu_r^{\rm init})+12.9333L(\mu_s)+2.09022\ln{\frac{\mu_r^{\rm init}}{m_b}}, \\
c_{3,2}&=& 0.257472+0.632361L(\mu_r^{\rm init})+0.783834L^2(\mu_r^{\rm init})-0.391917L(\mu_s), \\
c_{4,0}&=& -14311.9-1283.62L(\mu_h)+22709.0L(\mu_r^{\rm init})+6618.69L(\mu_h)L(\mu_r^{\rm init})+832.64L^2(\mu_r^{\rm init}) \nonumber\\
&& +7824.62L^3(\mu_r^{\rm init})+8857.05L(\mu_s)-6618.69L(\mu_h)L(\mu_s)-11736.9 L(\mu_r^{\rm init})L(\mu_s)  \nonumber\\
&& +8102.97L^2(\mu_s)+939.94L(\mu_{us})+518.834\ln{\frac{\mu_h}{m_b}} -3971.21L(\mu_r^{\rm init})\ln{\frac{\mu_h}{m_b}}  \nonumber\\
&& +160.453L(\mu_s)\ln{\frac{\mu_h}{m_b}}+40.1133\ln^2{\frac{\mu_h}{m_b}} -4651.9\ln{\frac{\mu_r^{\rm init}}{m_b}}
-2276.25L(\mu_r^{\rm init})\ln{\frac{\mu_r^{\rm init}}{m_b}}  \nonumber\\
&&-2647.48\ln{\frac{\mu_h}{m_b}}\ln{\frac{\mu_r^{\rm init}}{m_b}}-758.751\ln^2{\frac{\mu_r^{\rm init}}{m_b}}
-7097.75\ln{\frac{\mu_s}{m_b}}-511.158\ln^2{\frac{\mu_s}{m_b}}  \nonumber\\
&& +7586.97\ln{\alpha_s(\mu_s)} +3429.11\ln{\frac{\mu_s}{m_b}} \ln{\alpha_s(\mu_s)}-2727.7\ln^2{\alpha_s(\mu_s)}, \\
c_{4,1}&=& 1458.98-2785.37L(\mu_r^{\rm init})-401.133L(\mu_h)L(\mu_r^{\rm init})-957.856L^2(\mu_r^{\rm init})-1422.66L^3(\mu_r^{\rm init})  \nonumber\\
&& -212.745L(\mu_s)+401.133L(\mu_h)L(\mu_s)+2133.99L(\mu_r^{\rm init})L(\mu_s)-534.305L^2(\mu_s)  \nonumber\\
&& +97.0847\ln{\frac{\mu_h}{m_b}}+240.68L(\mu_r^{\rm init})\ln{\frac{\mu_h}{m_b}}+400.066\ln{\frac{\mu_r^{\rm init}}{m_b}}
+275.909L(\mu_r^{\rm init})\ln{\frac{\mu_r^{\rm init}}{m_b}}  \nonumber\\
&&+160.453\ln{\frac{\mu_h}{m_b}}\ln{\frac{\mu_r^{\rm init}}{m_b}}+91.9698\ln^2{\frac{\mu_r^{\rm init}}{m_b}}+133.615\ln{\frac{\mu_s}{m_b}}
-80.2265\ln^2{\frac{\mu_s}{m_b}},\\
c_{4,2}&=& -13.3227+120.457L(\mu_r^{\rm init})+89.7734L^2(\mu_r^{\rm init})+86.2217L^3(\mu_r^{\rm init})-9.85995L(\mu_s)  \nonumber\\
&& -129.333L(\mu_r^{\rm init})L(\mu_s)+12.9333L^2(\mu_s)-5.92731\ln{\frac{\mu_r^{\rm init}}{m_b}}
-8.36089L(\mu_r^{\rm init})\ln{\frac{\mu_r^{\rm init}}{m_b}}  \nonumber\\
&& -2.78696\ln^2{\frac{\mu_r^{\rm init}}{m_b}}, \\
c_{4,3}&=& 0.265871-1.71648L(\mu_r^{\rm init})-2.10787L^2(\mu_r^{\rm init})-1.74185L^3(\mu_r^{\rm init})+0.320593L(\mu_s) \nonumber\\
&& +2.61278L(\mu_r^{\rm init})L(\mu_s)-0.261278L^2(\mu_s),
\end{eqnarray}
\end{widetext}
where $\mu_h$, $\mu_s$ and $\mu_{\rm us}$ stand for the hard, the soft and the ultra-soft factorization scales, respectively. $L(\mu_r^{\rm init})$, $L(\mu_h)$, $L(\mu_s)$ and $L(\mu_{\rm us})$ are corresponding to taking $x=\mu_r^{\rm init},\mu_h,\mu_s,\mu_{\rm us}$ for $L(x)$, respectively.

The conformal coefficients $r_{i,0}(i=1,2,3,4)$ read (in unit of GeV),
\begin{eqnarray}
r_{1,0}&=& 0.0734844,\\
r_{2,0}&=& -3.33129,\\
r_{3,0}&=& 80.6951+60.1699L(\mu_h)-60.1699\ln\frac{\mu_h}{m_b}, \\
r_{4,0}&=& 7600.61-1283.62L(\mu_h)+4102.54L(\mu_s) \nonumber\\
&& +1634.33L^2(\mu_s)+939.94L(\mu_{us}) \nonumber\\
&& +160.453L(\mu_s)\ln\frac{\mu_h}{m_b}+2120.73\ln\frac{\mu_h}{m_b} \nonumber\\
&& +40.1133\ln^2\frac{\mu_h}{m_b}-4893.09\ln\frac{\mu_s}{m_b} \nonumber\\
&& -1834.9\ln^2\frac{\mu_s}{m_b}+7586.97\ln{\alpha_s(\mu_s)} \nonumber\\
&& +3429.11\ln\frac{\mu_s}{m_b}\ln{\alpha_s(\mu_s)} \nonumber\\
&& - 2727.7\ln^2{\alpha_s(\mu_s)}.
\end{eqnarray}
As required, these equations show that the conformal coefficients are free of initial scale dependence. The PMC scales $Q_i(i=1,2,3)$ with full initial scale and factorization scale dependence for each perturbative order read
\begin{widetext}
\begin{eqnarray}
\ln\frac{Q_1^2}{(\mu_r^{\rm init})^2} &=& -0.806755-2L(\mu_r^{\rm init})-\Big(1.32611-4L(\mu_s)\Big){\beta_0}a_s(\mu_r^{\rm init})
          +a_s^2(\mu_r^{\rm init}) \Big[\Big(10.1-2.65222L(\mu_r^{\rm init}) \nonumber\\
          && -8L(\mu_s)+8L(\mu_r^{\rm init})L(\mu_s)-4L^2(\mu_s)\Big) \beta_0^2-\Big(1.65764-5L(\mu_s)\Big)\beta_1\Big], \label{fac1} \\
\ln\frac{Q_2^2}{(\mu_r^{\rm init})^2} &=& -0.0020727-1.76471L(\mu_r^{\rm init})-0.235294\ln\frac{\mu_r^{\rm init}}{m_b}
          +a_s(\mu_r^{\rm init})\beta_0\Big[-2.06665+0.998408L(\mu_r^{\rm init}) \nonumber\\
          && -0.33218L^2(\mu_r^{\rm init})+3.84375L(\mu_s)-\Big(0.998408-0.66436L(\mu_r^{\rm init})\Big)\ln\frac{\mu_r^{\rm init}}{m_b}-0.33218\ln^2\frac{\mu_r^{\rm init}}{m_b} \Big],  \label{fac2} \\
\ln\frac{Q_3^2}{(\mu_r^{\rm init})^2} &=& \Big[5.77145-3.50367L(\mu_r^{\rm init})-2L(\mu_h)L(\mu_r^{\rm init})-0.671837L(\mu_s)+2L(\mu_h)L(\mu_s)-1.6L^2(\mu_s) \nonumber\\
          &&+\Big(0.484053+1.2L(\mu_r^{\rm init})+0.8\ln\frac{\mu_r^{\rm init}}{m_b}\Big)
\ln\frac{\mu_h}{m_b}+0.821428\ln\frac{\mu_r^{\rm init}}{m_b}+0.66619\ln\frac{\mu_s}{m_b} \nonumber\\
          &&-0.4\ln^2\frac{\mu_s}{m_b}\Big] / \Big(1.34112 +L(\mu_h)-\ln\frac{\mu_h}{m_b}\Big).  \label{fac3}
\end{eqnarray}
\end{widetext}

As a minor point, we have found that there are some typos for the general coefficients $r_{4,j}$ with $j=(0,1,2)$ at the four-loop level, i.e. Eqs.(39b-39d) of Ref.\cite{Brodsky:2013vpa} (they are correct for $n=1$) should be corrected as
\begin{widetext}
\begin{eqnarray}
r_{4,2} &=& \frac{1}{32(n+1)(n+2)T_F^3} \left[2T_F^2 c_{2,1}(79C_A+66C_F)-9\left(\frac{4(3+2n)}{n+1}T_F c_{3,2}(5C_A+3C_F)-33c_{4,3}C_A-4T_F c_{4,2}\right) \right], \\
r_{4,1} &=& \frac{1}{64(n+2)T_F^3} \Bigg[4 T_F^2 c_{2,1} (-397C_A C_F-118C_A^2-126C_F^2) + 48 T_F^2 c_{3,1} (5C_A+3C_F) \nonumber\\
&&+\frac{12 T_F c_{3,2}}{n+1} C_A \big((152n+173)C_A+33(4n+5)C_F \big)-33C_A \left(33c_{4,3}C_A+8T_F c_{4,2}\right)-48T_F^2 c_{4,1} \Bigg], \\
r_{4,0} &=& c_{4,0}+\frac{1}{64T_F^3}\Big[2T_F^2 c_{2,1} C_A (1208C_A C_F-287C_A^2+924C_F^2)-48T_F^2 c_{3,1} C_A (7C_A+11C_F) \nonumber\\
&&-2904 T_F c_{3,2} C_A^2 C_F+176 T_F^2 c_{4,1} C_A-1848 T_F c_{3,2} C_A^3 +484 T_F c_{4,2} C_A^2+1331 c_{4,3} C_A^3 \Big].
\end{eqnarray}
\end{widetext}

\end{document}